\documentclass[10pt,twoside]{hsqcd}
\usepackage{epsf,amsmath}
\usepackage{graphicx}
\usepackage{cite}

\begin{document}

\title{Experimental Summary of HSQCD2005
\thanks{presented at
HSQCD2005,  20 - 24th September, St. Petersburg.}
}

\author{\underline{J\"org Gayler} \\ \\      
DESY \\
}

\maketitle

\begin{abstract}
\noindent A summary of experimental contributions to HSQCD 2005 is given.
This includes results from the four HERA experiments H1,  ZEUS, HERMES
and HERA-B, as well as results from the Tevatron (D0), the 
CERN hyperon beam (WA98), from RHIC (PHENIX),  and IHEP (SVD).
I present a short survey of the points appearing most relevant to me.
%From these contributions I select the points which appeared
%most important to me.

\end{abstract}

%\markright{}

%\renewcommand{\@evenhead}

%\markboth{\large \sl \underline{J. Gayler}
%\hspace*{2cm} HSQCD 2005} {\large \sl \hspace*{1cm}
%HSQCD 2005 PROCEEDINGS}

\section {Introduction}

This summary is the write-up of 
 a quick and spontaneous tour through the  experimental
results presented at the conference
\footnote{In fact the speaker was found by tossing a coin
by Leonid Gladilin and myself during a coffee break.}, rather than a deep discussion of the 
experimental situation of QCD in general.

Among the accelerators, which presently provide results, HERA
 concentrates most on topics related to QCD, which is at the heart of interests
 of the HSQCD organisers at Gatchina. And,
 in spiritu loci, many
   analyses presented at this conference consider
  the  limitations of DGLAP evolution, ask
  whether BFKL evolution might be more appropriate. 
But there is progress in many other respects as well,
 e.g. parton density analyses
get more precise and more general, similarly, the data on charm and beauty
 production in $ep$ interactions improve rapidly,
 the spin structure of the proton is studied in more and more detail,
 RHIC continues to provide interesting results on hadronic high density states.
 Although not all experiments,
  which produce relevant results in context of QCD were represented,
 a wide spectrum of
  interesting results was displayed.  
      
   QCD is generally discussed    
   in the frame of the standard model. 
   I will therefore begin with 
  searches for physics beyond it.

  \section{Searches Beyond the Standard Model}
  \label{sec:bsm}
 
   D. South~\cite{south} reported on 
      searches at HERA for  physics beyond the standard model (SM) 
      which was not really shaken yet,
     although in $e^+p$ the H1 collaboration observes
     more events than expected 
     with isolated leptons and large missing
     transverse momentum and large $p_t^X$ ($ > 25$ GeV),
     the transverse momentum of the summed hadronic final state.
     However in this kinematic range the statistics is still rather small, 
     with 15 events with isolated electrons or muons for a SM expectation
     of $4.6 \pm 0.8$ events. This excess is not confirmed by the
     ZEUS collaboration and is not observed in $e^-p$ data.
     
     It is certainly very important to take as much data
     as possible, but it might be
     difficult to get full clarity in the remaining HERA 
     running time.   
     The same is true for the multi-lepton final states, where 
     6 multi-electron states with a di-electron invariant mass above 100 GeV
     were recorded by H1, 
      3 $ee$ events and 3 $eee$ events with SM expectations of 
      $0.44 \pm 0.10$ and $0.29 \pm 0.06$ events, respectively.

\section{Inclusive $ep$ scattering}
\label{sec:incl}

   With increasing statistics, HERA enters a phase of
    mature electro-weak physics analyses.
   The available NC and CC data 
    give the opportunity to demonstrate nicely the 
    unification of electro-weak physics (see ~\cite{ri}).
    The  absence of right handed weak currents could be demonstrated
    by the the dependence of the CC cross section on the polarisation
    of the incident $e^{\pm}$ beam (see Figs. in~\cite{ri}).
    The propagator mass in CC reactions is measured
    with a precision of about 2 GeV
    and found to be consistent  with the time like $W$ mass
    determined in $e^+ e^-$ and $\bar {p} p$. 
    Finally, first results are presented for light quark-$Z^0$
     couplings~\cite{Aktas:2005iv}
    which supplement the LEP and Tevatron measurements.
    
    Since its beginning, HERA is the world supplier for information
    on  the proton structure at low $x$. After the discovery of the
    rise of $F_2$ towards small $x$~\cite{Abt:1993cb}, this rise is
    described with continuously increasing precision in terms of
    parton density functions (pdfs) (see~\cite{jg}).
    Now the HERA collaborations provide pQCD analyses of their
    NC and CC data with the aim to include as little other data as possible,
    thereby avoiding additional systematic uncertainties.
    In particular, due to the sensitivity of
     the CC reaction $e^+p \rightarrow \bar {\nu} X$
    to  the valence $d$~quark distribution,  pdf fits without
     inclusion of $ed$ scattering data are possible,
    thereby avoiding nuclear corrections.
    The ZEUS collaboration provided also a pdf fit~\cite{Chekanov:2005nn}
    using jet data,
    in addition to the inclusive data, thereby exploiting 
    their sensitivity  to the gluon density via the boson-gluon
    fusion interaction.
     The pdf fits recently provided by the HERA collaborations are consistent
    with the results of the global fitters~\cite{Martin:2001es}. 
    
      Still increasing precision
     can be expected at small $x$ 
     from the final analysis of the full
    HERA I data including
    results at large $y$ which are sensitive to the longitudinal structure
    function $F_L$ and therefore
    to the gluon density.
    In addition new techniques are explored to improve the   
    NC measurements at very high $x$~\cite{jg}.
    Thus, further progress in the determination of parton densities
    can be expected  in future.    
                    
\section{Jets}
\label{sec:jets}

      Complementary to inclusive $ep$ scattering (previous section), 
      the convolution of proton pdfs with the hard
     interaction processes  can be studied in detail
     by the 
     observation of jets in deep inelastic scattering (DIS)
     and photoproduction~\cite{sutton}.
      The high precision HERA data at large $Q^2$ and large transverse momenta
     provide, besides detailed comparisons with     
     pQCD calculations, results on the strong coupling constant $\alpha_s$
     which turn out to be consistent and competitive with results 
     from inclusive   
     $ep$ or $e^+e^-$ data. 
     As mentioned already above, the
     inclusion of $\gamma p$ and DIS
      jet data in pdf fits improves the determination of
     the gluon density in the proton at medium and large $x$, where the
     inclusive NC data have only little sensitivity.
     In most of the phase space, the experimental uncertainties of the 
     available jet data are smaller than the theoretical ones. 
     
     Extra complications appear in the theoretical description of
      dijet production in DIS at low $Q^2$ and large $E_T^2$. Here     
    a resolved virtual photon contribution
    is expected,  where a partonic fraction
     of the virtual photon  enters the hard interaction. This 
      is supported by the data which are not described 
      in this region by NLO calculations with only
      direct photon interactions. 
     
     Such a resolved component improves also the description of 
     ``forward jet" data~\cite{dobur}.
      By requiring $E_T^{jet} \sim Q^2$ and jet angles
     as close to the proton beam as possible with the H1 and ZEUS calorimeters,
     a phase space is selected, where the data might discriminate the 
     calculations
     based on DGLAP evolution or CCFM or  BFKL evolutions.
     Less jet production is obtained in the  $k_t$ ordered DGLAP evolution
     than in unordered emissions like in the CDM model,
     which might be close to BFKL expectations, or the CCFM
     approach (angular ordering).
     Indeed, the DGLAP based models  produce too small cross sections at
     small $x_{Bjorken}$ compared to the measurements of H1 and ZEUS.
      However the best descriptions are obtained
     in conceptually quite different models  like  CDM
     with unordered emissions, and the DGLAP based RAPGAP model,
      where the strict
     $k_t$ ordering is broken by a resolved photon component.  
     Interesting to note that for a more special event selection, requiring
     besides the forward jet two additional central jets, the resolved
     RAPGAP model also fails, in contrast to CDM~\cite{Aktas:2005up}.
     
     Impressive data on inclusive single jet production in $p\bar {p}$ were
      presented by the D0 collaboration~\cite{obrant}. The $p_t$
      distributions agree   
      beautifully with pQCD in NLO. Angular correlations demonstrate
      very well   that there is no need for higher orders. However the
      large effect of the experimental uncertainty of the jet energy
      scale  on the cross 
         section measurement
      allows no conclusions on pdfs to be drawn.
      
\section{Inclusive Diffraction}
\label{sec:diff}
 
      The observation of diffractive phenomena in DIS at HERA gives 
       hope to reach, in terms 
       of QCD, 
       full understanding  of
             the Regge exchange processes, which dominate the strong
              interactions processes at high energies.
       
       So far, the efforts to describe the colour singlet exchange in 
       diffractive
       interactions by pdfs are partially successful, as
       diffractive inclusive data, diffractive jet production and diffractive
       charm production can indeed be described in a common 
       approach~\cite{dodonov}.
       These pdfs, however, can not be naively applied to diffractive
       $p\bar {p}$ data, not even to diffractive photoproduction of jets.
       In the latter case, the predictions, which are based on the pdfs 
        deduced from inclusive diffraction, are above the data by about
        a factor two. This may indicate that in photoproduction, as in
         $p\bar {p}$ interactions, the gap signature of the diffractive
         interaction is more often
          destroyed than in DIS
           by additional 
          parton-parton interactions.
        More results are expected from HERA and the hadron colliders which     
          may  help to
          describe diffractive phenomena coherently.    

\section{Vector Mesons and DVCS}
\label{sec:VM}

   New results were presented on $J/\psi$ photo- and electro production
   at low $t$, on
     $J/\psi$ production in DIS at high $t$,
     as well as high $t$ $\rho^0$ photoproduction, and on $\phi$  production
     in DIS~\cite{berger}. 
     
     Photo and electro-production of 
     vector mesons (VM) is an interesting testing ground for QCD due to 
     the interplay of soft processes described by Regge theory and hard    
     interactions described by pQCD. For the latter different hard
     scales may be involved, $Q^2$, the VM mass, the momentum
     transfer $t$. The presented results show that whenever hard scales are   
     involved, the data
     can not be described by soft pomeron exchange.    
     Vector mesons are particularly suited for detailed QCD
     calculations as the measured decay angular distributions provide
     information on the helicity structure of the interaction.
      Such an analysis was presented
     for $\gamma p \rightarrow \rho p$ at high $t$~\cite{berger}.
     
     $J/\psi$ production in 920 GeV proton-carbon interactions where
     presented by the HERA-B collaboration which extend the available
     $x_{Feynman}$ distributions far into the target region~\cite{egorytchev}
     (see also the discussion \cite{ryzhinsky}).   
     
   Results on Deeply Virtual Compton Scattering (DVCS, 
     $ep \rightarrow \gamma p$) were presented
     by the HERMES~\cite{tytgat},  
      H1 and ZEUS collaborations~\cite{berger}. 
     The reaction $ep \rightarrow \gamma p$,
     which, in contrast 
     to VM production,
      does not require knowledge of a non-perturbative wave function,
     gives in principle access to generalised
     non collinear parton densities. 
      HERMES obtains  DVCS amplitudes by  
      measuring asymmetries due to the interference of Bethe Heitler (BH)
      $\gamma $ emissions from electrons with the DVCS process.  
      There are already first results on $e$-beam-spin and on beam-charge
       asymmetries
      and on longitudinal and transverse target spin asymmetries, both on 
      $p$ and $d$ targets. 
        More data are expected with improved control of the
       elasticity of the final state with the new Recoil Detector of
       HERMES.
       
       H1 and ZEUS have not yet exploited the  BH-DVCS
       interference,
    which vanishes when integrated over the phase space of the measurements,
      but provide DVCS cross section measurements which are
       described by NLO pQCD calculations and various colour dipole
       models.  
     The measured energy dependence of the cross section is characteristic
   of processes where a hard scale is involved. 
   Precise measurements of the $t$ dependence allow the theoretical predictions
    to be normalised.
     It remains to hope that the future data will distinguish 
       better between the different theoretical approaches.

\section{Charm and Beauty Production}
\label{sec:charm}
      Heavy flavour production in electro and photoproduction is expected to be
       well described by pQCD due to the large scale given by the heavy quark
       mass. It was therefore surprising, that in spite 
       of this advantage, beauty $(b)$ production cross sections were
        notoriously
       above the pQCD NLO predictions~\cite{bell}. 
       This is still the case, but it is interesting to note
       that the available more inclusive measurements with large 
       acceptance, which use $b$ identification by life time tags, are
       consistent with the predictions.
       
       The contributions of charm and beauty to the inclusive proton
       structure function $F_2$ are substantial, about 20\% to 30\% for
       charm and about 0.3\% to 3\% for beauty, 
      in the covered
       kinematic region~\cite{bell,jg}.
       
       Fragmentation properties of $c$ quarks were studied by H1
       in DIS and with more statistics in $\gamma p$
       by   ZEUS~\cite{gladilin}.  
        By and large, the findings are consistent with
        $e^+e^-$ results, supporting ``universality"
       of $c$~quark fragmentation.
        
    Results on  hadroproduction of open beauty near threshold were presented
    by HERA-B~\cite{egorytchev} which are reasonably consistent with
    Fermilab data and pQCD calculations which include NNLO soft gluon
     corrections~\cite{Kidonakis:2004qe}.

\section{Spin Physics}
\label{sec:spin}

\subsection{HERMES results}

 The HERMES experiment~\cite{tytgat} produces a wealth of results
 on the many measurable effects of the proton spin in 
 electron nucleon scattering when besides the scattered electron one
 hadron is detected. 
  In particular the effects of
 the nucleon spin polarisation transverse to the electron beam
 direction (``transversity") lead to
   sophisticated analyses which distinguish between the effect of the
   quark polarisation on the $p_t$ transverse to it
    in fragmentation (Collins effect),
   and the effect of the struck quarks $p_t$ in the polarised nucleon
   on the $p_t$ in fragmentation (Sivers effect), which gives access to the 
   quark angular momentum.
   
   For  $\pi^-$ the Collins moments are negative but larger in absolute
    than expected.   
   The Sivers moments  show a striking difference between $\pi^+$ and 
   $\pi^-$, the latter being consistent with zero, the first definitely
   positive which points to the valence $u$ content of
   the $\pi ^+$. 
   
   The expected future data and further  
   theoretical analyses might eventually lead to a clear understanding
   of the  proton spin\footnote{
   The HERMES results on DVCS are mentioned in section~\ref{sec:VM}.}.

\subsection{Hyperon polarisation}
\label{sec:hyp}

   Many results exist on the polarisation of hyperons produced in 
   hadron beams, without clear theoretical understanding~\cite{siebert}.
   In particular, the results obtained by WA98 in the 340 GeV $\Sigma^-$
   beam 
   on the polarisation of inclusively
   produced  
   $\Lambda $ and $\Sigma^-$ baryons originally came as a surprise
     with a  large unexpected positive
   polarisation of the $\Lambda$ 's (normal to the production plane).
    Meanwhile explanations exist in terms of recombination of a quark
    with a strange di-quark~\cite{Kubo:2005af}. However no coherent     
    theoretical approach
    to the observed detailed polarisation phenomena exists yet.
    ``Food for theorists!" (W.H. Siebert~\cite{siebert})

\section{Elliptic Flow in Nucleus-Nucleus Collisions}

RHIC data show a strong azimuthal anisotropy of the particle flow
in Au Au collisions at $\sqrt{s_{NN}} = 200$ GeV~\cite{taranenko}.
The particle flow is increased in the direction transverse to the reaction
plane.  It is understood to be sensitive to  the initial geometric overlap
of the colliding nuclei as well as the later expansion driven by the
initial pressure. The effect is seen for all final state hadrons and 
the results suggest~\cite{taranenko}
 early formation of the anisotropy in a still partonic
phase. 

To clarify the mechanism, it is particularly
 interesting to measure the
energy flow for direct photons. As not strongly  interacting, direct photons
penetrate through hadronic clouds and are known to be not suppressed in heavy
ion collisions~\cite{Adler:2005ig} in contrast to hadrons.
 With present statistics the data on prompt photons are
not yet conclusive~\cite{Adler:2005rg}.

\section{Pentaquarks}
\label{sec:penta}

  A new result  of a pentaquark search was reported~\cite{popovsvd}
 from the SVD-2 detector using the 70 GeV proton beam at IHEP.
 In the reaction $pA \rightarrow pK_s^0 + X$
  a bump is seen at the $pK_s^0$ mass 
  $M = 1526 \pm 2 ({\rm stat.}) \pm 3$ (syst.) MeV
  with a width $\Gamma < 24$ MeV, which can be interpreted
  as the signal of $uudd\bar{s}$ pentaquark $\Theta^+$.    
  The estimated significance is 5.6 $\sigma$.

   The many negative and positive evidences for pentaquarks were 
   discussed in a review by V. Popov~\cite{popov}.
   L. Gladilin~\cite{gladilin} reported on the results from HERA which are
   not yet clear,
   with ZEUS seeing a $\Theta^+$ signal in contrast to H1 and H1 seeing
   a candidate for a charmed pentaquark $\Theta_c$ at a mass of 3100 GeV
   which is not confirmed by ZEUS and other experiments.  
   Negative results were reported for
   $\Xi_{3/2}^{--}$~\cite{gladilin,siebert}.
   
   Also other non standard hadronic states were discussed, namely
   possible tensor glue balls with masses around 2000 MeV~\cite{mateev} 

  The great time of rapid pentaquark discoveries seems to be over, but there are 
  still many interesting signals which need final confirmation or
  other explanations. 
   
\section{FAIR, a forthcoming facility for QCD physics}

  The approved {\bf F}acility for {\bf A}ntiproton and {\bf I}on {\bf R}esearch
 (FAIR)
  at GSI, Darmstadt (Germany) is expected to provide in 2013
  very intensive beams
  (e.g. primary beams of $10^{10}/s \; ^{238} {\rm U}^{73+}$ up to
  35 GeV/u and $3 \times 10^{13}/s$ protons of 30 GeV)~\cite{ritman}.
  Besides various radioactive secondary beams, a $\bar {p}$ beam
  will be available  for the planned experiment PANDA
  which foresees a rich program on
   hadron spectroscopy, in particular
  for searches of glue balls or exotic hadrons and detailed studies
  of charmonia and production of open charm. It is seen as an advantage of
   $p\bar {p}$ compared to $e^+e^-$ interactions that the formation of states
   is not restricted to the virtual photon quantum numbers $1^{--}$.     
   
   Thus, FAIR will be after 2013
 an interesting facility for hadron physics 
   in the low and medium energy range.

\section{Conclusion}

%\begin{itemize}

%\item 
  \noindent
There is still intensive experimental activity addressing QCD related
questions.

%\item
 \noindent
 HERA, the Tevatron, RHIC will provide important information before the
 LHC will produce results.

%\item
 \noindent
We experimentalists have not yet presented results which need BFKL
beyond any doubt, however some final state analyses prefer non $k_t$ 
ordered gluon emissions.

%\item
\noindent
 At next HSQCD we will know considerably more.
 
% \end{itemize}

\section*{Acknowledgements}  
I thank all the organisers, in particular Victor Kim and Lev Lipatov
for  a stimulating conference with very lively discussions in  the sessions
and the breaks. 
I thank Emmanuelle~Perez
for reading the draft and helpful comments.


\begin{thebibliography}{0}

%\cite{south}
\bibitem{south}
%\cite{South:2006dc}
%\bibitem{South:2006dc}
D.~M.~South  [H1 Collaboration], these proceedings;
%``Searches for new physics at HERA,''
arXiv:hep-ex/0602028.
%%CITATION = HEP-EX 0602028;%%

\bibitem{ri}
   Yongdok Ri, these proceedings; ibid. J. Gayler;
%\cite{Aktas:2005ju}
%\bibitem{Aktas:2005ju}
A.~Aktas {\it et al.}  [H1 Collaboration],
%``First measurement of charged current cross sections at HERA with
%longitudinally polarised positrons,''
Phys.\ Lett.\ B {\bf 634}, 173 (2006)
[hep-ex/0512060];
%%CITATION = HEP-EX 0512060;%%
%\cite{unknown:2006da}
%\bibitem{unknown:2006da}
  [ZEUS Collaboration],
%``Measurement of high-Q**2 deep inelastic scattering cross sections with a
%longitudinally polarised positron beam at HERA,''
hep-ex/0602026.
%%CITATION = HEP-EX 0602026;%% 

%\cite{Aktas:2005iv}
\bibitem{Aktas:2005iv}
A.~Aktas {\it et al.}  [H1 Collaboration],
%``A determination of electroweak parameters at HERA,''
Phys.\ Lett.\ B {\bf 632} (2006) 35.
%[arXiv:hep-ex/0507080].
%%CITATION = HEP-EX 0507080;%%



%\cite{Abt:1993cb}
\bibitem{Abt:1993cb}
I.~Abt {\it et al.}  [H1 Collaboration],
%``Measurement of the proton structure function F2 (x, Q**2) in the low x region
%at HERA,''
Nucl.\ Phys.\ B {\bf 407} (1993) 515;
%%CITATION = NUPHA,B407,515;%%
%\cite{Derrick:1993ft}
%\bibitem{Derrick:1993ft}
M.~Derrick {\it et al.}  [ZEUS Collaboration],
%``Measurement of the proton structure function F2 in e p scattering at HERA,''
Phys.\ Lett.\ B {\bf 316} (1993) 412.
%%CITATION = PHLTA,B316,412;%%

\bibitem{jg} J. Gayler, these proceedings, hep-ex/0603037.

\bibitem{Chekanov:2005nn}
S.~Chekanov {\it et al.}  [ZEUS Collaboration],
%``An NLO QCD analysis of inclusive cross-section and jet-production data  from
%the ZEUS experiment,''
Eur.\ Phys.\ J.\ C {\bf 42} (2005) 1.
%[arXiv:hep-ph/0503274].
%%CITATION = HEP-PH 0503274;%%

%\cite{Martin:2001es}
\bibitem{Martin:2001es}
A.~D.~Martin, R.~G.~Roberts, W.~J.~Stirling and R.~S.~Thorne,
%``MRST2001: Partons and alpha(s) from precise deep inelastic scattering  and
%Tevatron jet data,''
Eur.\ Phys.\ J.\ C {\bf 23} (2002) 73.
%[arXiv:hep-ph/0110215];
%%CITATION = HEP-PH 0110215;%%
%\cite{Pumplin:2002vw}
%\bibitem{Pumplin:2002vw}
J.~Pumplin, D.~R.~Stump, J.~Huston, H.~L.~Lai, P.~Nadolsky and W.~K.~Tung,
%``New generation of parton distributions with uncertainties from global  QCD
%analysis,''
JHEP {\bf 0207} (2002) 012.
%[arXiv:hep-ph/0201195].
%%CITATION = HEP-PH 0201195;%%

\bibitem{sutton}  
 M.~Sutton, these proceedings.

\bibitem{dobur} 
 D.~ Dobur, these proceedings.
 
%\cite{Aktas:2005up}
\bibitem{Aktas:2005up}
A.~Aktas {\it et al.}  [H1 Collaboration],
%``Forward jet production in deep inelastic scattering at HERA,''
hep-ex/0508055.
%%CITATION = HEP-EX 0508055;%% 

\bibitem{obrant}
G.~Obrant, these proceedings.


\bibitem{dodonov}
V.~Dodonov, these proceedings.

\bibitem{berger}
N.~Berger, these proceedings.

\bibitem{egorytchev}
V.~Egorytchev, these proceedings.

\bibitem{ryzhinsky}
M.~Ryzhinsky, these proceedings.

\bibitem{tytgat}
M.~Tytgat, these proceedings.

\bibitem{bell}
M.~Bell, these proceedings.

\bibitem{gladilin}
L.~Gladilin, these proceedings.

%\cite{Kidonakis:2004qe}
\bibitem{Kidonakis:2004qe}
N.~Kidonakis and R.~Vogt,
%``Threshold corrections in bottom and charm quark hadroproduction at
%next-to-next-to-leading order,''
Eur.\ Phys.\ J.\ C {\bf 36} (2004) 201.
%[hep-ph/0401056].
%%CITATION = HEP-PH 0401056;%%

\bibitem{siebert}
W.H.~Siebert, these proceedings.

%\cite{Kubo:2005af}
\bibitem{Kubo:2005af}
K.~I.~Kubo and K.~Suzuki,
%``Spin polarization of hyperons in hadron-hadron inclusive collisions,''
hep-ph/0505179.
%%CITATION = HEP-PH 0505179;%%

\bibitem{taranenko}
A.~Taranenko, these proceedings.

%\cite{Adler:2005ig}
\bibitem{Adler:2005ig}
S.~S.~Adler {\it et al.}  [PHENIX Collaboration],
%``Centrality dependence of direct photon production in s(NN)**(1/2) =  200-GeV
%Au + Au collisions,''
Phys.\ Rev.\ Lett.\  {\bf 94} (2005) 232301.
%[arXiv:nucl-ex/0503003].
%%CITATION = NUCL-EX 0503003;%%

%\cite{Adler:2005rg}
\bibitem{Adler:2005rg}
S.~S.~Adler {\it et al.}  [PHENIX Collaboration],
%``Measurement of identified pi0 and inclusive photon v(2) and implication to
%the direct photon production in s(NN)**(1/2) = 200-GeV Au + Au collisions,''
Phys.\ Rev.\ Lett.\  {\bf 96} (2006) 032302.
%[arXiv:nucl-ex/0508019].
%%CITATION = NUCL-EX 0508019;%%

%\cite{popovsvd}
\bibitem{popovsvd} V.~Popov, these proceedings;
%\cite{Kubarovsky:2005sk}
%\bibitem{Kubarovsky:2005sk}
A.~Kubarovsky and V.~Popov  [SVD Collaboration],
%``Observation of narrow baryon resonance in p K0(S) mode in p A interactions at
%70-GeV/c with SVD-2 setup,''
hep-ex/0510006.
%%CITATION = HEP-EX 0510006;%%

%\cite{popov}
\bibitem{popov}
 V.~Popov, these proceedings.

%\cite{mateev}
\bibitem{mateev}
 M. Mateev, these proceedings.

%\cite{ritman}
\bibitem{ritman}
 J.~Ritman, these proceedings.


\end{thebibliography}
\end{document}